\input{aipcheck.tex}
\documentclass[
   final
  ]
  {aipproc}

\usepackage{epsf,epsfig}

\newcommand{\be}{\begin{equation}}
\newcommand{\ee}{\end{equation}}
\newcommand{\bea}{\begin{eqnarray}}
\newcommand{\eea}{\end{eqnarray}}
\newcommand{\nn}{\nonumber}
\newcommand{\ba}{\begin{array}}
\newcommand{\ea}{\end{array}}
\newcommand{\bi}{\begin{itemize}}
\newcommand{\ei}{\end{itemize}}

\newcommand{\N}{\chi^0}
\newcommand{\C}{\chi^\pm}
\newcommand{\lsim}{
\mathrel{\hbox{\rlap{\hbox{\lower4pt\hbox{$\sim$}}}\hbox{$<$}}}}
\newcommand{\gsim}{
\mathrel{\hbox{\rlap{\hbox{\lower4pt\hbox{$\sim$}}}\hbox{$>$}}}}

\newcommand{\half}{\frac{1}{2}}

\renewcommand{\slash}{\displaystyle{\not}}

\layoutstyle{6x9}

\begin{document}


\title
{Singlet Extensions of the MSSM}


\author{Vernon Barger}{
address={Department of Physics, University of Wisconsin, Madison, WI 53706 }
}
\author{Paul Langacker}{
address={ School of Natural Sciences, Institute for Advanced Study, Einstein Drive, Princeton, NJ 08540}
}
\author{Gabe Shaughnessy}{
address={Department of Physics, University of Wisconsin, Madison, WI 53706 }
}

\begin{abstract}
\noindent
The Higgs sector of the MSSM may be extended to solve the $\mu$ problem by the addition of a gauge singlet scalar field.  We consider the consequences of the singlet on the Higgs and neutralino states compositions.  We discuss the potential for discovering Higgs bosons of singlet models and observing the unique multilepton signatures of the extended neutralino sector.
\end{abstract}
\maketitle


\section{Introduction}
The Minimal Supersymmetric Standard Model (MSSM) is a leading candidate for beyond the standard model (SM) physics.  The motivation for the MSSM is extensive and includes solutions to the gauge hierarchy problem, the quadratic divergence in the Higgs boson mass, gauge coupling unification, and a viable dark matter candidate.   

However, there are challenges to the MSSM.  Fine tuning is required to obtain the correct value for the $Z$ boson mass.  A light scalar top and Higgs boson are required if electroweak baryogenesis is to account for the total baryon asymmetry in the universe \cite{Carena:1997gx}.  Additionally, the allowed region of parameter space in constrained versions of the MSSM that is consistent with the relic density of neutralinos, the popular dark matter candidate, is limited, see, for example, Ref. \cite{Ellis:2005mb}.

The MSSM has an important theoretical problem associated with the Higgsino mixing parameter, $\mu$, which is the only massive parameter that is supersymmetry conserving.  The value of $\mu$ sets the scale of the electroweak symmetry breaking in the MSSM and is thus required to be at the electroweak (EW) or TeV scale, though a priori it can have any value.  

Supersymmetric models with an additional singlet Higgs field address this fine-tuning problem of the MSSM by promoting the $\mu$ parameter to a dynamical field whose vacuum expectation value $\langle S\rangle$ and coupling $\lambda$ determine the effective $\mu$-parameter,
\be
\mu_{\rm eff} = \lambda \langle S\rangle.
\ee
Depending on the symmetry imposed on the theory, a variety of singlet extended models (xMSSM) may be realized: see Table \ref{tbl:models}.  The models we focus on include the Next-to-Minimal Supersymmetric SM (NMSSM) \cite{NMSSM}, the Nearly-Minimal Supersymmetric SM (nMSSM) \cite{nMSSM,Menon:2004wv}, and the $U(1)'$-extended MSSM (UMSSM) \cite{UMSSM}, as detailed in Table \ref{tbl:models} with the respective symmetries.  A Secluded $U(1)'$-extended MSSM (sMSSM) \cite{sMSSM,Han:2004yd} contains three singlets in addition to the standard UMSSM Higgs singlet; this model is equivalent to the nMSSM in the limit that the additional singlet vevs are large, and the trilinear singlet coupling, $\lambda_s$, is small \cite{Barger:2006dh}.   The nMSSM and sMSSM will therefore be referred to together as the n/sMSSM.  The additional singlet state of the extended models gives additional Higgs bosons and neutralino states.  The number of Higgs and neutralino states in the various models are summarized in Table \ref{tbl:models}.

\begin{table}[ht]
\begin{tabular}{|c|ccccc|}
\hline
Model:& MSSM &NMSSM &nMSSM& UMSSM & sMSSM\\
\hline
Symmetry:  &--  &~~ $Z_3$    & $Z^R_5, Z^R_7$      & ~~$U(1)'$&$U(1)'$ \\\hline
Extra   &--         &       ~~${\kappa\over3} \hat S^3$    & $t_F  \hat S$& ~~-- &$\lambda_S S_1 S_2 S_3$ \\
superpotential term&   --      &      (cubic)    & (tadpole) & ~~-- & (trilinear secluded)\\\hline
 $\N_i$ &        	4	  &	~~~5	&  	5	 & ~~6 & 9\\
 $H^0_i$ &       	2	  &	~~~3	&  	3	 & ~~3 & 6\\
 $A^0_i$ &        	1	  &	~~~2	&  	2	 & ~~1 & 4\\
\hline
\end{tabular}
\caption{Symmetries associated with each model and their respective terms in the superpotential; the number of states in the neutralino and Higgs sectors are also given.  All models have two charginos, $\C_i$, and one charged Higgs boson, $H^\pm$.  We ignore possible CP violation in the Higgs sector.}
\label{tbl:models}
\end{table}

The additional CP-even and CP-odd Higgs boson, associated with the inclusion of the singlet field, yield interesting experimental consequences at colliders.  For recent reviews of these models including their typical Higgs mass spectra and dominant decay modes, see Ref. \cite{Barger:2006dh,Kraml:2006ga}.

The tree-level Higgs mass-squared matrices are found from the potential, $V$, which is a sum of the $F$-term, $D$-term and soft-terms in the lagrangian, as follows.
\bea
V_F &=& |\lambda H_u\cdot H_d+t_F+ \kappa S^2|^2 + |\lambda S|^2 \left(|H_d|^2+|H_u|^2 \right), \\
V_D &=& \frac{G^2}{8}\left( |H_d|^2-|H_u|^2 \right)^2+ \frac{g_{2}^2}{2} \left( |H_d|^2|H_u|^2-|H_u \cdot H_d|^2 \right),\\
 &+& {{g_{1'}}^2\over2}\left(Q_{H_d} |H_d|^2+Q_{H_u} |H_u|^2+Q_{S} |S|^2\right)^2\\ \nn
V_{\rm soft}&=&m_{d}^{2}|H_d|^2 + m_{u}^{2}|H_u|^2+ m_s^{2}|S|^2 + \left( A_s \lambda S H_u\cdot H_d + {\kappa \over 3} A_{\kappa} S^3+t_S  S + h.c. \right).
\label{eq:potential}
\eea
Here, the two Higgs doublets with hypercharge $Y=-1/2$ and $Y=+1/2$, respectively, are
\be
H_d = \left( \begin{array}{c} H_d^0 \\ H^- \end{array} \right), \qquad
H_u = \left( \begin{array}{c} H^+ \\ H_u^0 \end{array} \right).
\ee
and $H_u \cdot H_d = \epsilon_{ij} H_u^i H_d^j$.  For a particular model, the parameters in $V$ are understood to be turned-off appropriately
\bea
{\rm NMSSM}&:& g_{1'}=0,\  t_F = 0,\  t_S = 0,\nn\\
{\rm nMSSM}&:& g_{1'}=0,\ \kappa=0,\ A_\kappa = 0, \\
{\rm UMSSM}&:& t_F = 0,\ t_S = 0,\ \kappa = 0,\ A_\kappa = 0.\nn
\eea
The couplings $g_1,g_2$, and ${g_{1'}}$ are for the $U(1)_Y$, $SU(2)_L$, and $U(1)'$ gauge symmetries, respectively, and the parameter $G$ is defined as $G^2=g_1^2+g_2^2$.  
The NMSSM model-dependent parameters are $\kappa$ and $A_\kappa$ while the nMSSM parameters are $t_F$ and $t_S$.  The model dependence of the UMSSM is expressed by the $D$-term that has the $U(1)'$ charges of the Higgs fields, $Q_{H_d}, Q_{H_u}$ and $Q_S$. 

One loop radiative corrections to the Higgs mass can be large due to the large top quark Yukawa coupling.  At the one-loop level, the top and stop loops are the dominant contributions.  Gauge couplings in the UMSSM are small compared to the top quark Yukawa coupling so the one-loop gauge contributions can be dropped.  The model-dependent contributions do not affect the Higgs mass significantly at one-loop order.  Thus, the usual one-loop SUSY top and stop loops are universal in these models.  The one-loop corrections to the potential are derived from the Coleman-Weinberg potential.

\section{Higgs sector}

\begin{figure}[t]
\includegraphics[angle=-90,width=0.49\textwidth]{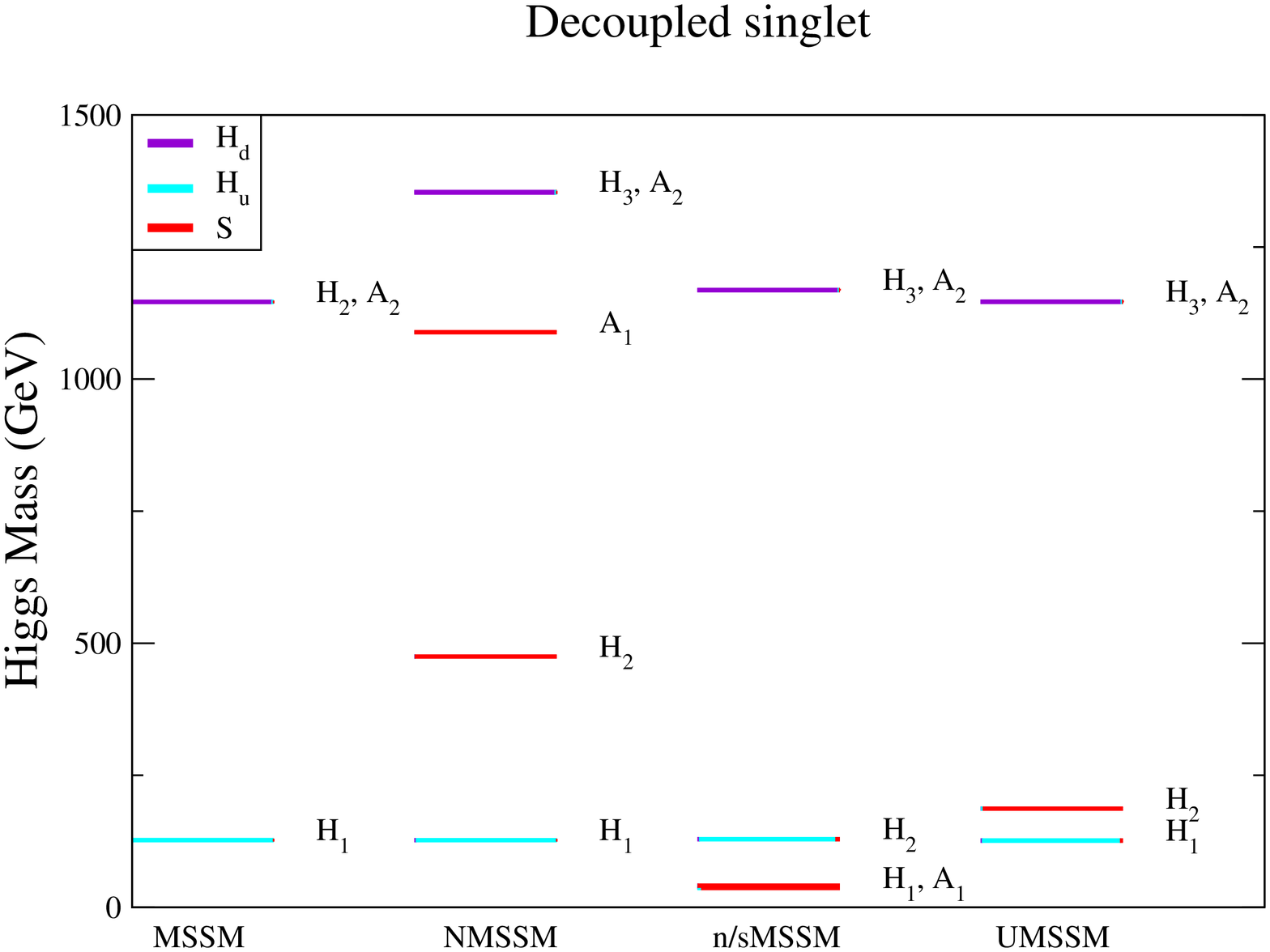}
\includegraphics[angle=-90,width=0.49\textwidth]{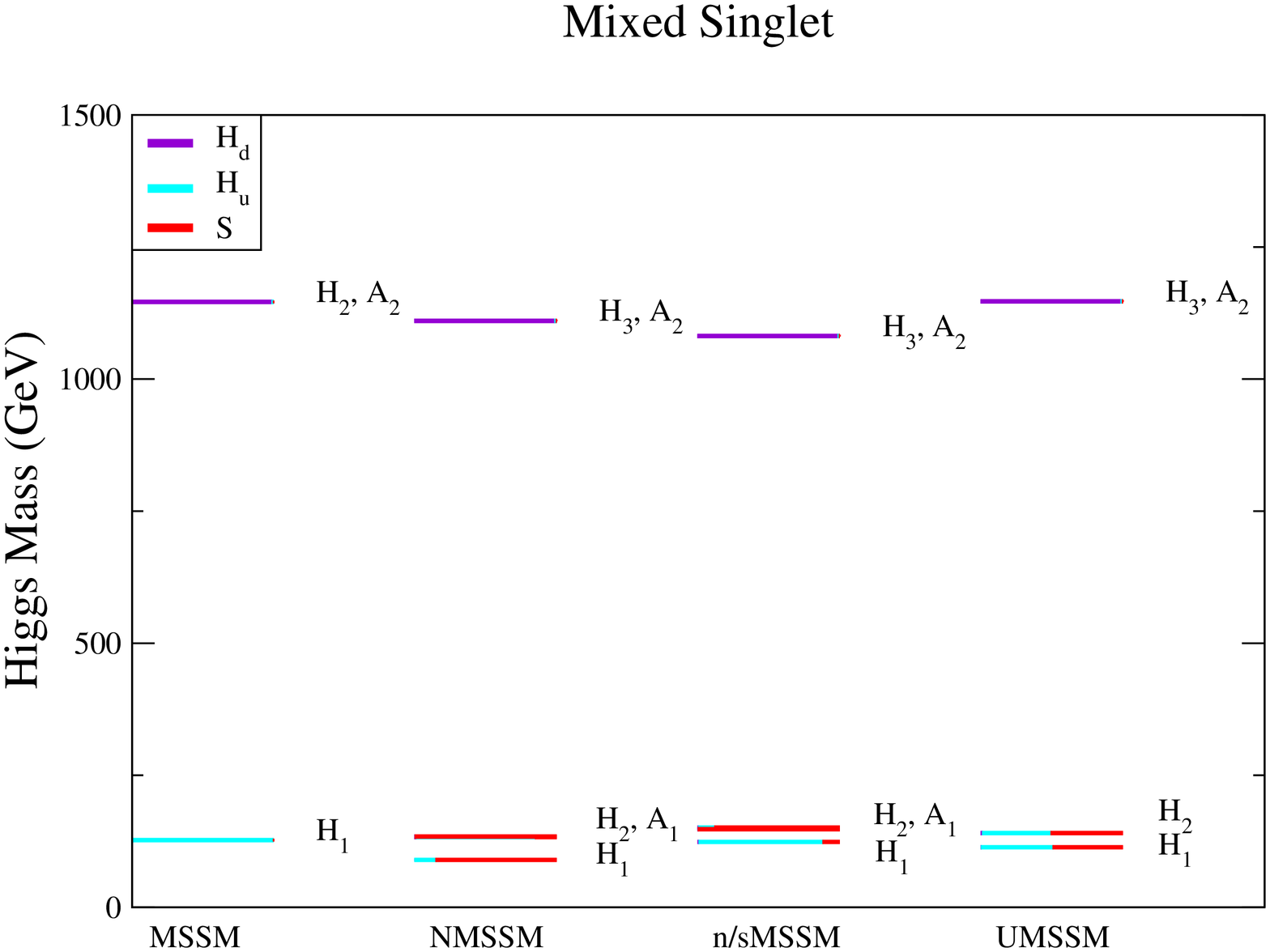}
\caption{Illustrative Higgs composition $(H_d, H_u, S)$ for the models in (a) a decoupled singlet scenario and (b) a strongly mixed singlet scenario.  In the decoupled scenario, the extended model has a spectrum similar to that of the MSSM, but contains an additional singlet Higgs that is heavy in the NMSSM and UMSSM and light in the n/sMSSM.  Figures from Ref. \cite{Barger:2006new}. }
\label{fig:illust}
\end{figure}

To illustrate the Higgs sector of the extended models in the cases in which the lightest Higgs is either decoupled or strongly mixed with the MSSM Higgs boson, we present in Fig. \ref{fig:illust} the neutral Higgs mass spectra for particular points in parameter space.

\begin{figure}[htbp]
\includegraphics[angle=-90,width=0.39\textwidth]{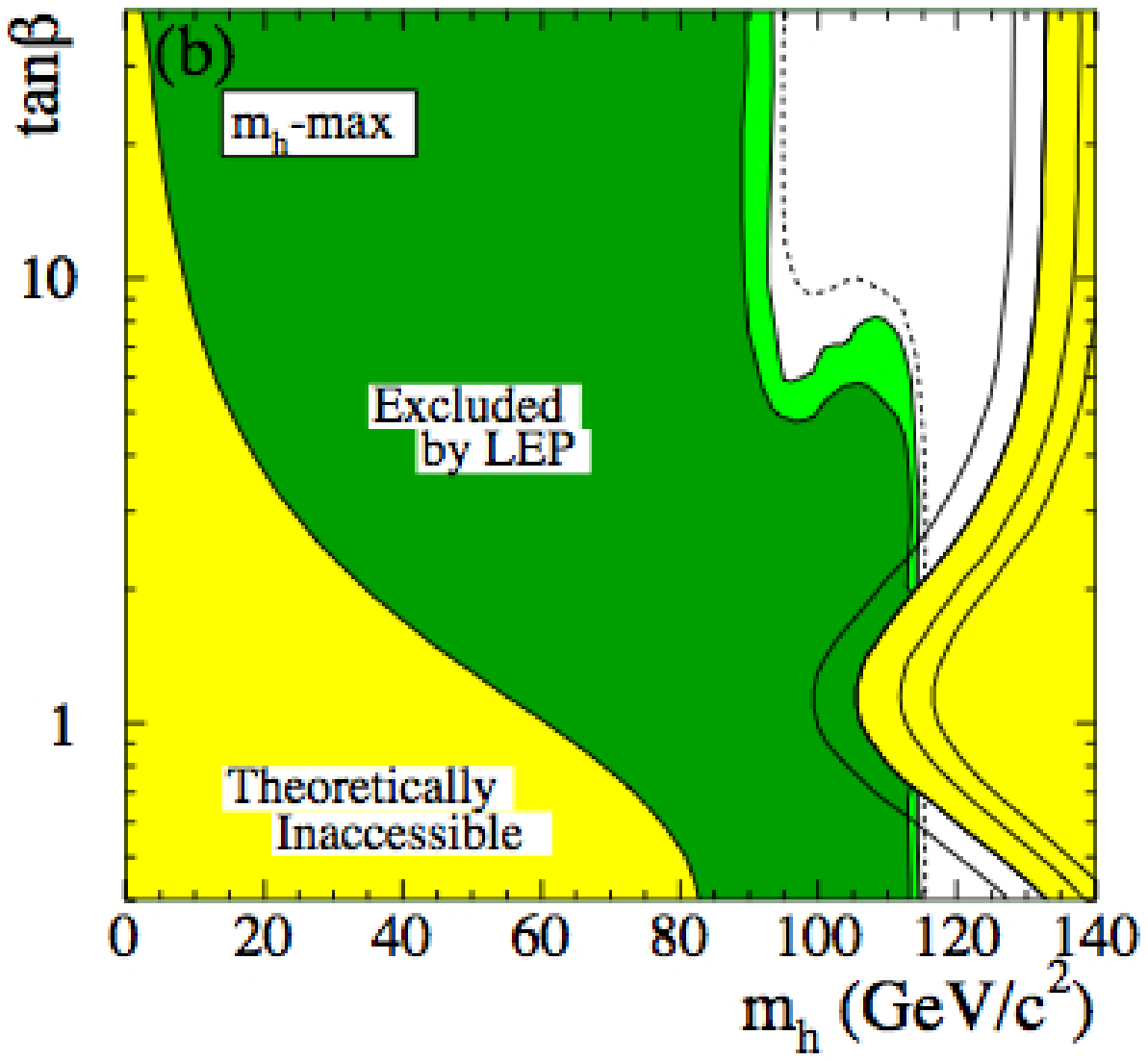}
\includegraphics[angle=-90,width=0.49\textwidth]{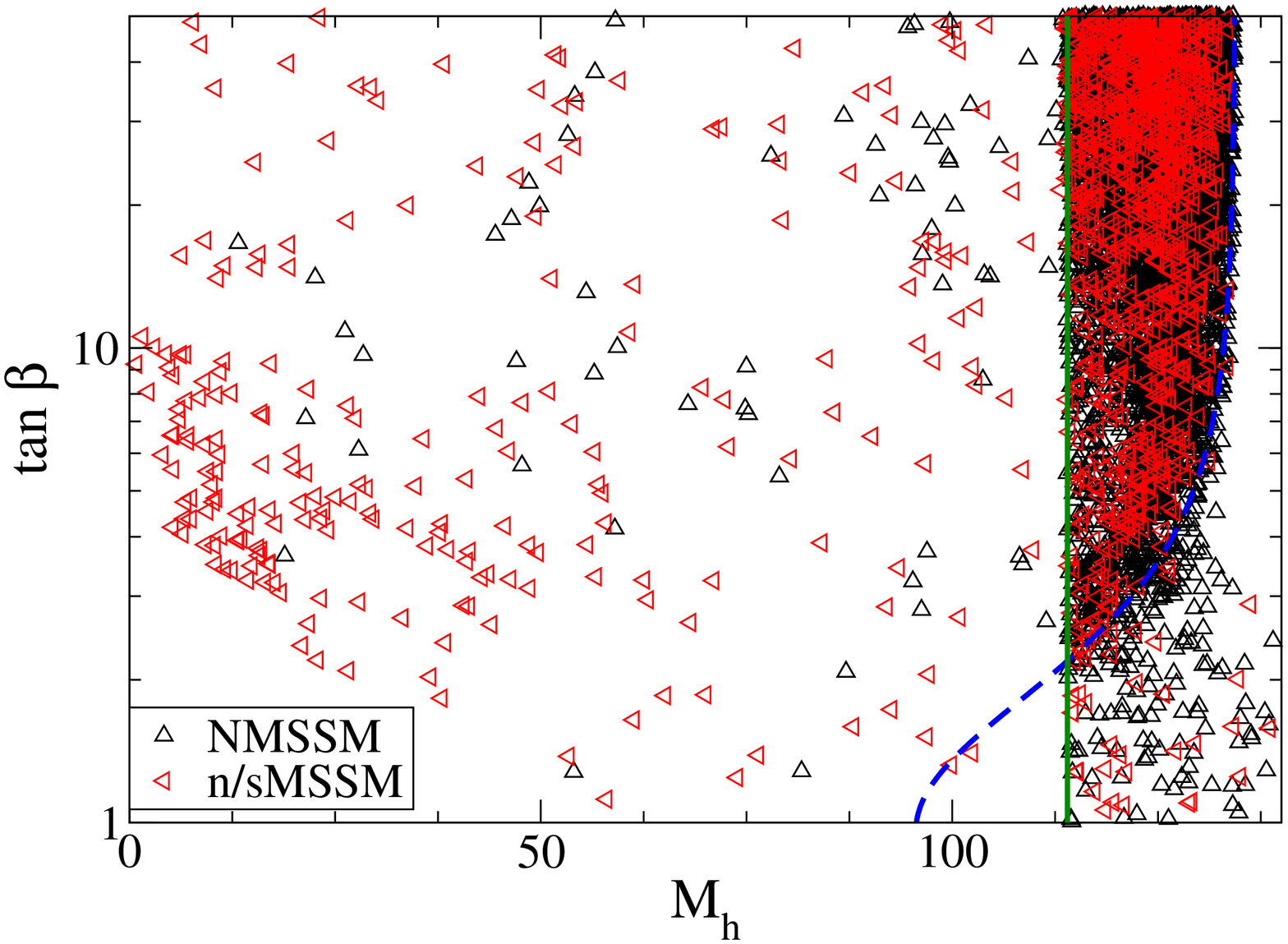}
\caption{(a) LEP exclusion region of the Higgs mass regions vs. $\tan \beta$.  The theoretically disallowed regions are shown in yellow.  (b) Singlet mixing allows Higgs masses below the SM LEP limit (shown as a vertical green line) and above the theoretical bound in the MSSM (shown as the blue dashed curve) due to singlet interactions. Figures from Refs. \cite{Sopczak:2006vn} and \cite{Barger:2006dh}.}
\label{fig:leplim}
\end{figure}

\begin{figure}[htbp]
\includegraphics[angle=-90,width=0.49\textwidth]{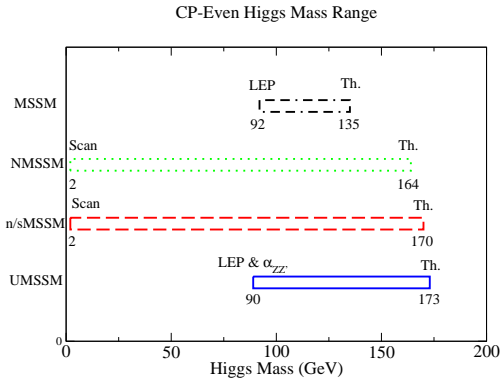}
\caption{Lightest CP-even Higgs mass range.  Figure from Ref. \cite{Barger:2006dh}.}
\label{fig:massrange}
\end{figure}

With sufficient mixing, the lightest Higgs boson can evade the current LEP bound on the SM Higgs mass in these models \cite{Barger:2006dh}.  This can be seen in Fig. \ref{fig:leplim} where the lightest Higgs boson can have masses that are inside the MSSM region excluded by LEP.  Alternatively, singlet interactions increase the lightest Higgs mass by ${\cal O}(\half \lambda^2 v^2\sin^2 2 \beta)$, allowing it to be in the theoretically excluded region in the MSSM for low $\tan \beta$ \footnote{Additional gauge interactions contribute to this increase with size ${\cal O}(g_{1'}^2 v^2(Q_{H_u}^2 \cos^2\beta+Q_{H_d}^2 \sin^2 \beta))$ in the UMSSM.}.  The lightest Higgs mass ranges for each model are shown in Fig. \ref{fig:massrange}.

\begin{figure}[t]
\includegraphics[angle=-90,width=0.45\textwidth]{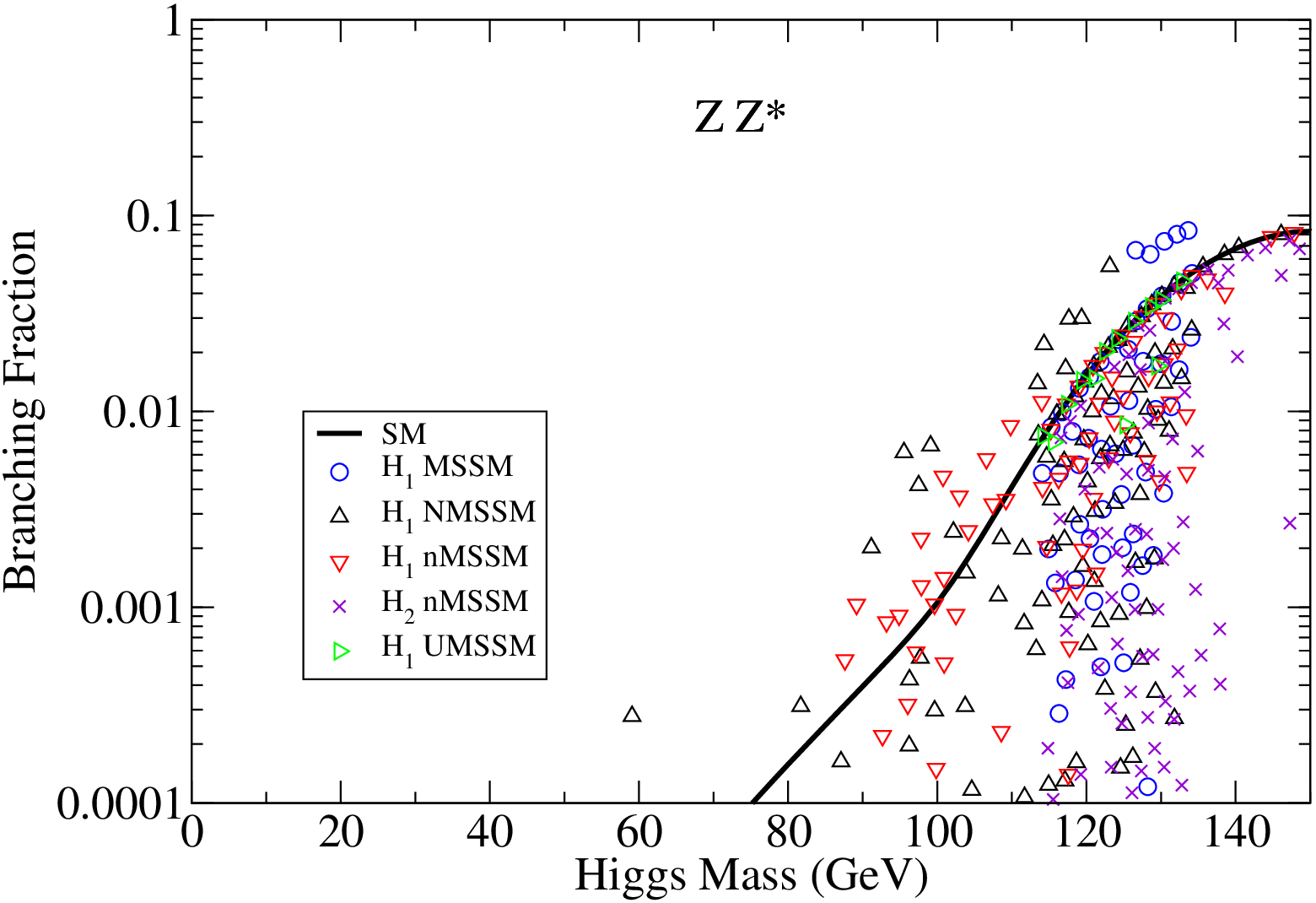}
\includegraphics[angle=-90,width=0.45\textwidth]{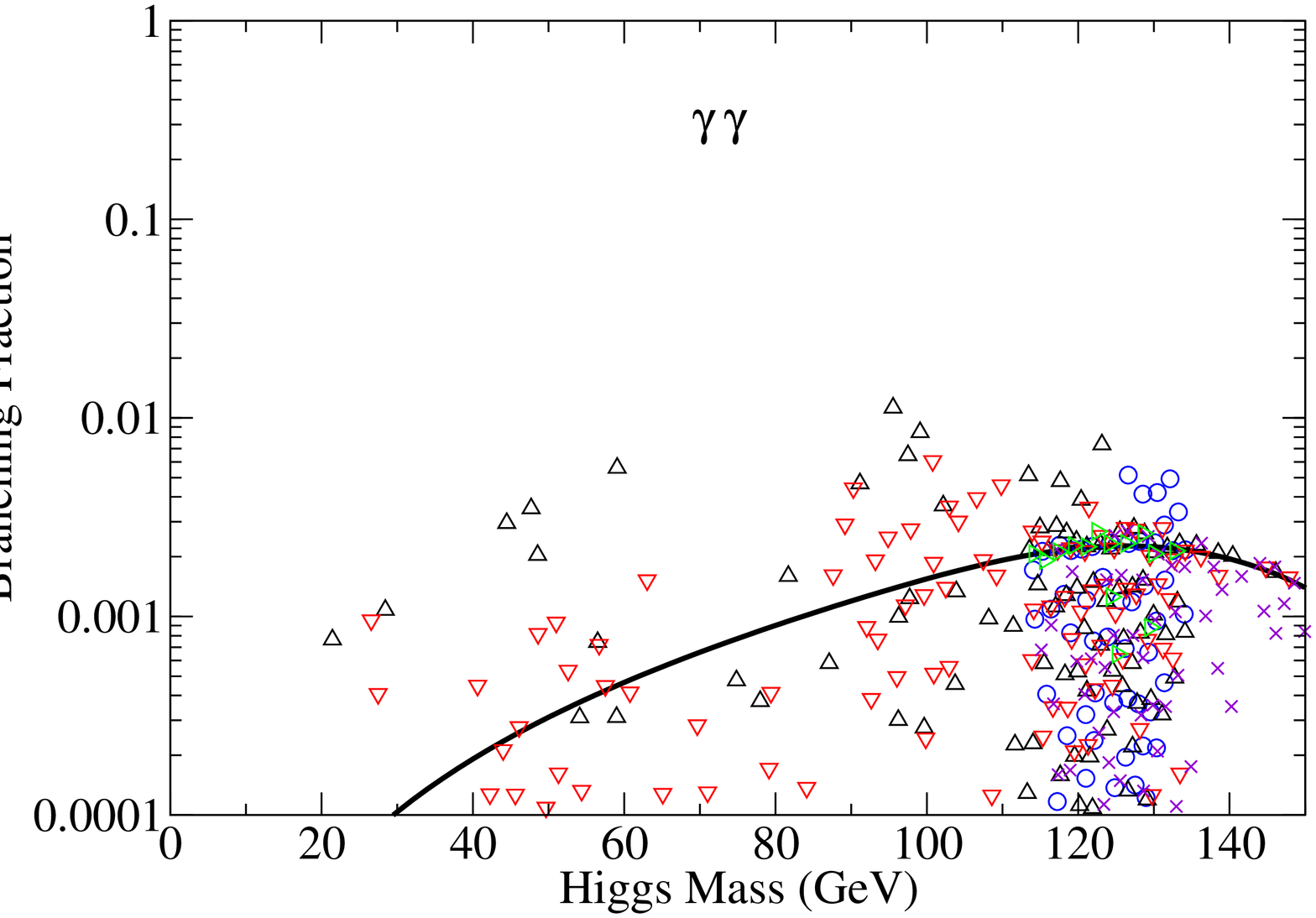}
\caption{Branching fractions to $Z$ boson pairs and photon pairs in the MSSM and extended-MSSM models. SM branching fractions are denoted by solid curves.  Figures from Ref. \cite{Barger:2006dh}.}
\label{fig:decays}
\end{figure}

A light decoupled $H_1$ is often accompanied by a MSSM-like $H_2$ with a mass in the 115-135 GeV range and MSSM strength couplings to SM fields.  Singlet mixing can strongly affect the observation of the Higgs boson at the LHC.  As seen in Fig. \ref{fig:decays}, the branching fractions for discovery channels of the Higgs boson in the SM can be suppressed significantly.  The couplings to gauge bosons are at most SM strength, and production rates are usually smaller than in the SM.

The most promising discovery channel over most of the Higgs mass range is the golden channel $H_i \to ZZ^* \to 4l$, since it has very low backgrounds.  This channel is expected to permit SM Higgs discovery for Higgs masses $120- 600$ GeV.  In extended models the signal is reduced by a factor of $\xi^2_{VVH_i}\times {Bf(H\to ZZ)\over Bf(h_{SM}\to ZZ)}$ compared to the SM, where $\xi_{VVH_i}$ is the $VVH_i$ coupling relative to the SM.  Therefore, it is possible that the Higgs in the extended models is missed via direct searches.

For light Higgs bosons ($m_H < 120$ GeV) the decay $H\to \gamma \gamma$ has the best significance.  Combining this mode with $H\to ZZ\to 4l$ yields a total significance above $5\sigma$ required for discovery for the lightest Higgs boson in the SM.  For some parameter points, the decay $H\to \gamma \gamma$ is enchanced due to a larger yukawa coupling or interference effects \cite{Barger:2006dh}.  In Fig. \ref{fig:decays}, we show the branching fractions of the Higgs states that have masses below 150 GeV to the promising modes $ZZ$ and $\gamma \gamma$.

\section{Neutralino sector}

\begin{figure}[t]
\includegraphics[angle=-90,width=0.49\textwidth]{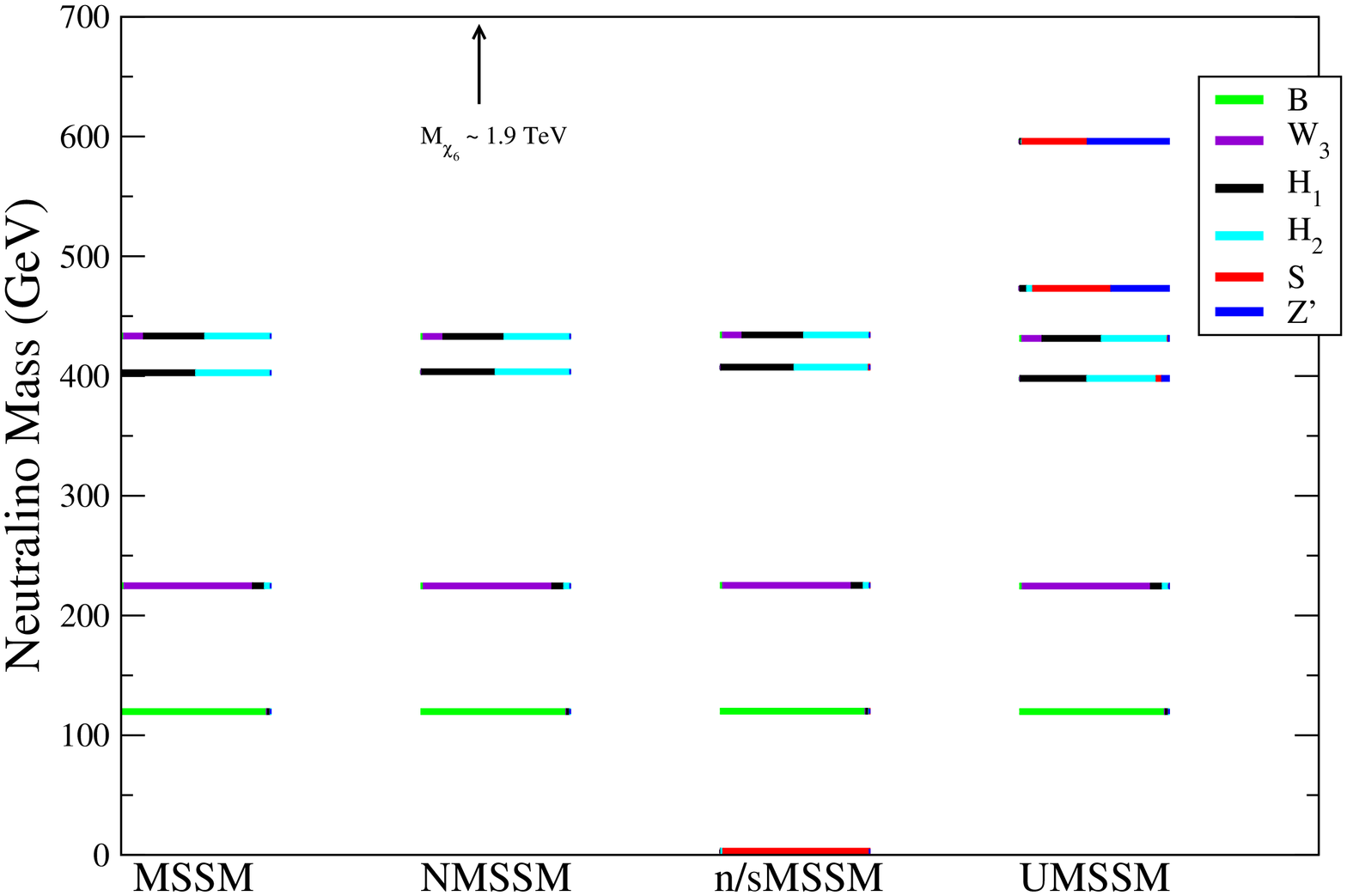}
\includegraphics[angle=-90,width=0.49\textwidth]{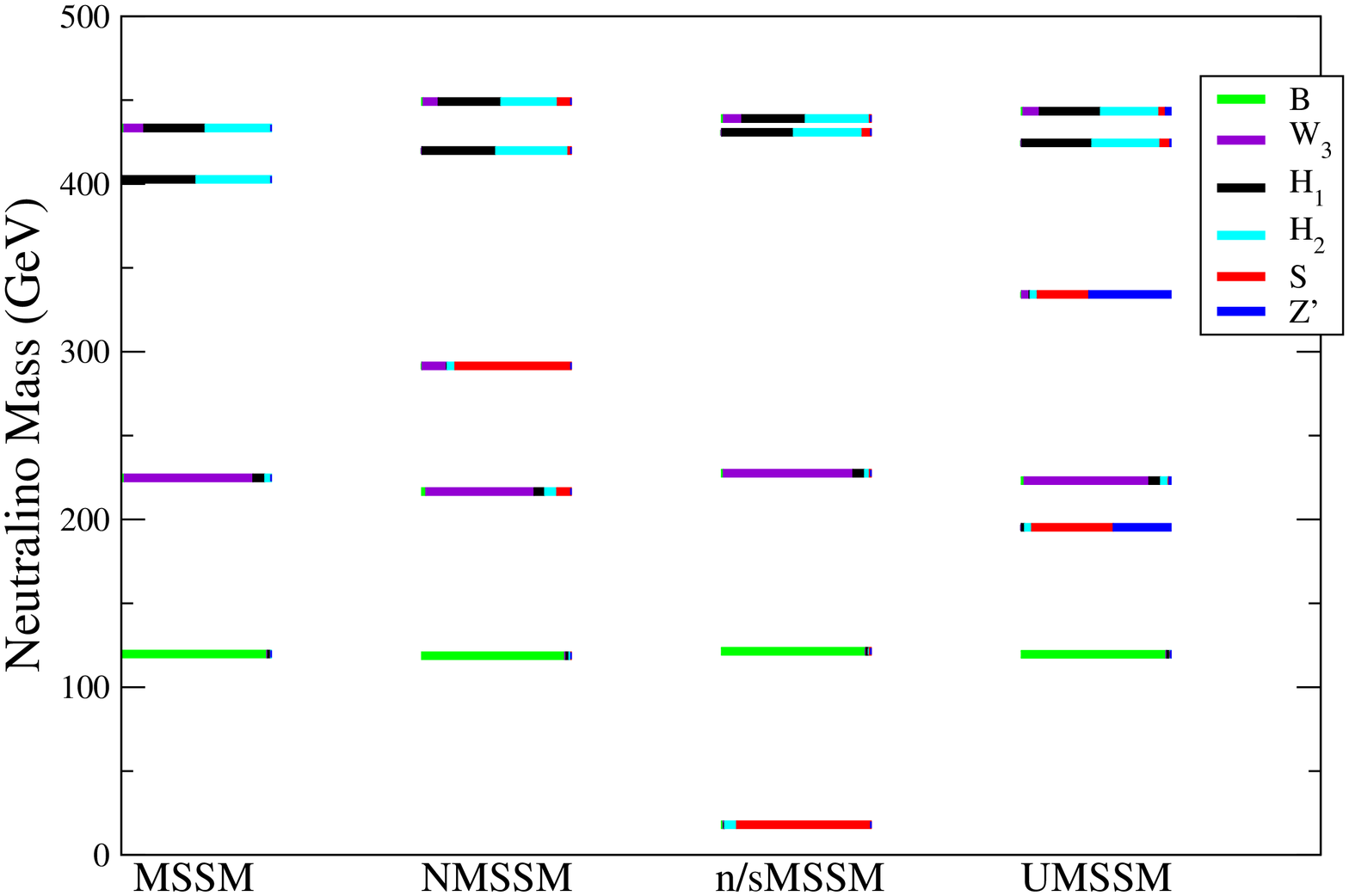}
\caption{Illustrative neutralino composition for the models in (a) a decoupled singlino scenario and (b) a strongly mixed singino scenario.  Here, the MSSM contains a light Bino and Wino and heavy Higgsinos.  The NMSSM has a similar spectrum, but contains an additional heavy neutralino, while the n/sMSSM has a very light extra neutralino.  The UMSSM has two additional neutralinos that can intermix; their masses are strongly dependent on the singlet Higgs charge under the $U(1)'$ symmetry and the corresponding gaugino mass value.  Figures from Ref. \cite{Barger:2006kt}.}
\label{fig:level-light}
\end{figure}

In the singlet extended models at least one new neutralino state beyond the MSSM exists.  The neutralino states can include four MSSM-like states and one nearly decoupled singlino state, or the singlino can significantly mix with the other states, as determined from the neutralino mass matrix
\bea
{\cal M}_{\N} = \left( \begin{array} {c c c c |c| c}
	M_1 	&	0&	{-g_1 v_d/ 2}&	{g_1 v_u / 2}&	0&  0\\
	0 	&M_2&	{g_2 v_d / 2}&	{-g_2 v_u / 2}&	0&  0\\
	{-g_1 v_d / 2} 	&	{g_2 v_d / 2}&	0&	-\mu_{\rm eff}&	-\mu_{\rm eff}v_u/s&  {g_{1'}} Q_{H_d} v_d\\
	{g_1 v_u / 2} 	&	{-g_2 v_u / 2}&	-\mu_{\rm eff}& 0&	-\mu_{\rm eff}v_d/s&  {g_{1'}} Q_{H_u} v_u\\
	\hline
	0&0&-\mu_{\rm eff} v_u/s&-\mu_{\rm eff} v_d/s&\sqrt 2 \kappa s&{g_{1'}} Q_{S} s\\
	\hline
	0&0&{g_{1'}} Q_{H_d} v_d&{g_{1'}} Q_{H_u} v_u&{g_{1'}} Q_{S} s& M_{1'}\\
	\end{array} \right).
	\label{eq:neutmass}
\eea
where $M_1$, $M_2$ and $M_{1'}$ are the gaugino masses of the $U(1)$, $SU(2)$ and $U(1)'$ gauge symmetries.  We assume gaugino mass unification, which constrains $M_{1'}=M_1={5 g_1^2\over3 g_2^2}M_2$ at low scales.  The resulting neutralino spectrum can be substantially altered with respect to the MSSM.  Figure \ref{fig:level-light} illustrates the neutralino spectrum and composition for a decoupled and mixed scenario of singlino (and $Z'$ino for the UMSSM) mixing.  Due to the shifts in the neutralino spectrum compared to the MSSM, the cascade decay chains may be substantially modified \cite{Barger:2006kt,Ellwanger:1997jj}.  In particular, excess trilepton and dilepton events can occur in models with a light singlino state.  

The neutralino in the n/sMSSM is very light, often below 50 GeV.  A very light neutralino in the n/sMSSM can allow a light stau that is not the LSP.  In the other singlet models, a very light singlino is less natural but can be achieved in the NMSSM with a very small value of $\kappa$, as the $\kappa \to 0$ limit corresponds to the n/sMSSM.  

Multilepton events such as a 5 lepton or 7 lepton signature are possible in extended models.  Chargino decays are indirectly affected via their decays to a lighter neutralino state.  The number of neutralino states lighter than the chargino and their modified compositions alter the chargino branching fractions.  This is typically found in the n/sMSSM, where the chargino can decay to an MSSM like $\N_2$ and a singlino $\N_1$, yielding a 5 lepton signal.  Additionally, the extra step in a neutralino decay can allow a 7 lepton final state.  Other models can also exhibit this behavior, but less naturally.

\begin{figure}[htbp]
\includegraphics[angle=-90,width=0.49\textwidth]{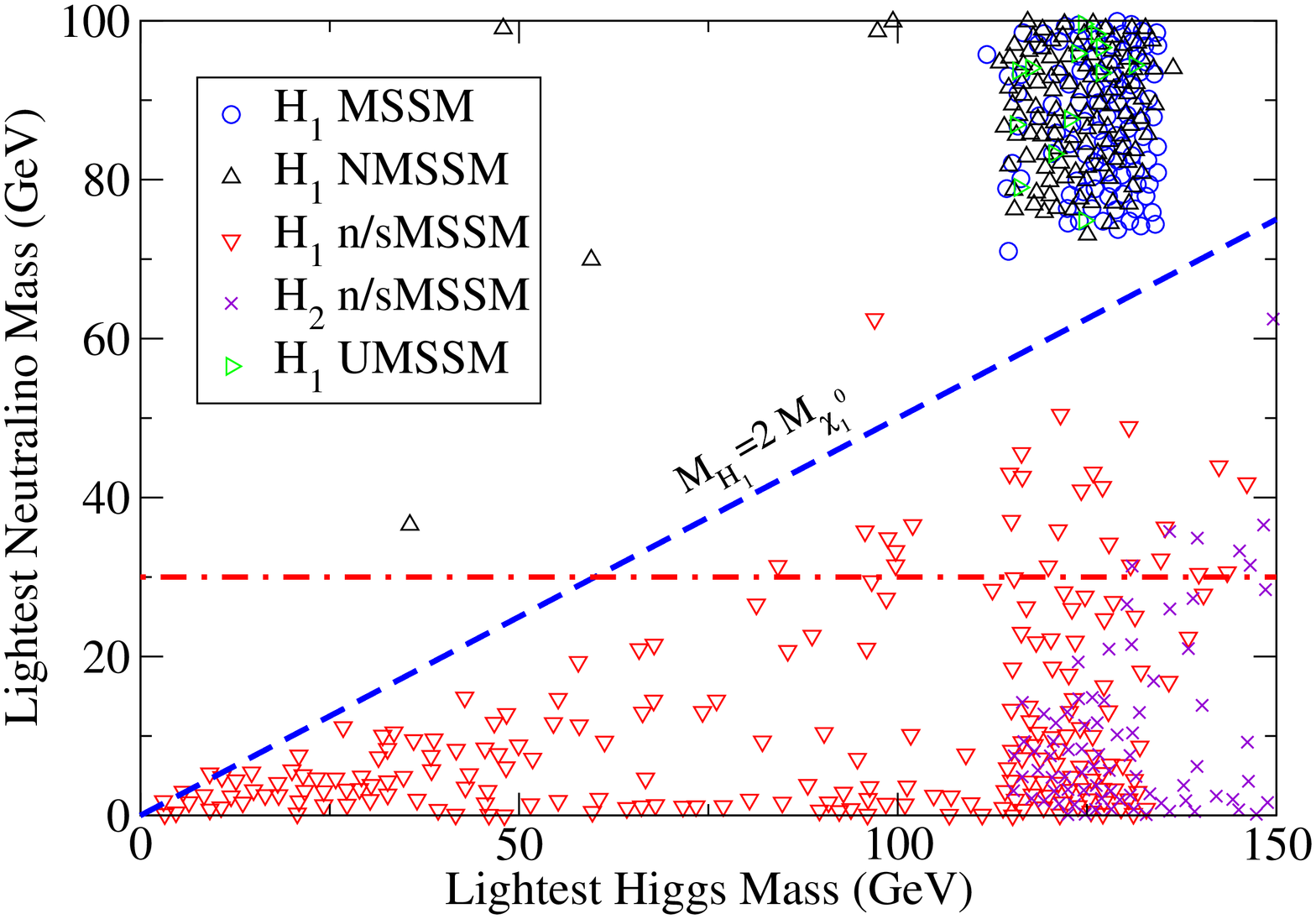}
\includegraphics[angle=-90,width=0.39\textwidth]{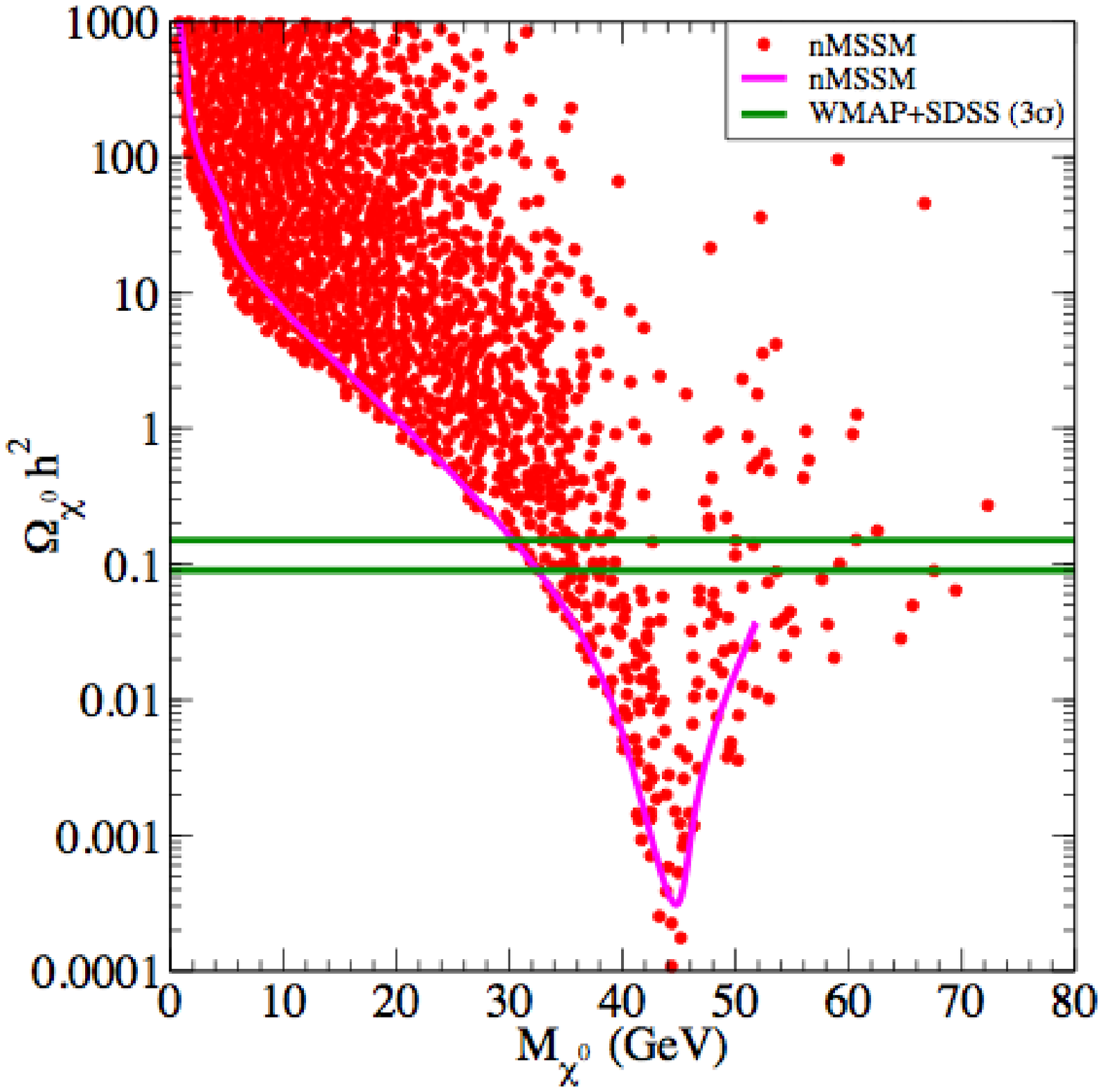}
\caption{(a) Higgs mass vs. lightest neutralino mass.  The kinematic region allowing decays $H_1 \to \N_1 \N_1$ is given below the blue dashed line.  These decays can be dominant, preventing traditional modes from being an effective means to discover the Higgs.  Figure taken from Ref. \cite{Barger:2006dh}. (b) Relic density vs. lightest neutralino mass in the n/sMSSM.  A lower bound on the lightest neutralino mass in this model can be placed at 30 GeV.  Figures taken from Refs. \cite{Barger:2006dh} and \cite{Barger:2005hb}.}
\label{fig:kinematicdec}
\end{figure}

In some cases the neutralino can be light enough to spoil the chances for direct Higgs discovery.  The Higgs boson may have a dominant invisible decay to stable neutralinos that are undetected except as missing transverse energy, $\slash E_T$.  When the $H\to \N_1 \N_1$ decay channel  is open, the Higgs is generally invisible \footnote{There are some corners of parameter space which allow $H_1\to A_1 A_1$ with the $A_1$ mass below the threshold for decays to bottom pairs \cite{Dermisek:2005ar}.}.  As seen in Fig. \ref{fig:kinematicdec}, this invisible decay is usually kinematically inaccessible for the MSSM, NMSSM, and UMSSM due to the lower limit on $m_{\N_1}$ of 53 GeV \cite{Barger:2006dh}.

Invisible decays are often dominant in the n/sMSSM where the lightest neutralino mass is typically lighter than 50 GeV \cite{Menon:2004wv,Barger:2006dh,Barger:2006kt,Barger:2005hb}.  Therefore, traditional searches for the discovery of $H_1$ is unlikely for some parameter regions of the n/sMSSM.  However, indirect discovery of an invisibly decaying Higgs is possible in WBF and in $Z$-Higgstrahlung at the LHC \cite{Eboli:2000ze,Davoudiasl:2004aj} with jet azimuthal correlations and $p_T$ distributions or via the $Z$ recoil spectrum at the ILC.

The relic density of dark matter provides a lower limit on the mass of the lightest neutralino in the n/sMSSM since the relic density becomes too large.  As seen in Fig. \ref{fig:kinematicdec}(b), the mass bound is $> 30$ GeV in this model \cite{Barger:2005hb} assuming only the $Z$ boson in the $s$-channel annihilation of the neutralinos.  If other annihilation processes contribute (or decays to a still lighter, almost decoupled, additional neutralino can occur, as in the sMSSM), the neutralino mass bound would be softened.

\section{Conclusions}

Higgs singlet extensions of the MSSM provide well motivated solutions to the $\mu$ problem.  Including an additional Higgs singlet increases the number of CP-even and CP-odd Higgs states and increases the number of associated neutralino states.  The extended models have interesting consequences in collider phenomenology.  Specifically, we find the following:

\bi

\item The lightest Higgs can be lighter than the LEP limit of $m_h > 114$ GeV due to reduced Higgs couplings to SM fields due to singlet-doublet mixing; the production rates of these Higgs states are often below the rates for the lightest MSSM Higgs boson.  

\item Direct observation of the lightest Higgs is favored for the MSSM, NMSSM and UMSSM.  In the n/sMSSM, the traditional discovery modes can be spoiled by the decay to invisible states such as neutralinos.  However, indirect observation of the Higgs can be employed for the n/sMSSM where invisible Higgs decays to neutralino pairs are often dominant.  

\item The extended models can have an approximately decoupled neutralino that is dominantly singlino, accompanied by an approximate MSSM spectrum of neutralino states.  The lightest neutralino is typically very light in the n/sMSSM, often below 50 GeV, and can affect the predicted number of multilepton events significantly.  The rate for $\N_{i\ge2} \C_1$ production increases since $\N_i$ is lighter than it would otherwise be in the MSSM.  The decoupled neutralinos in the NMSSM and UMSSM are typically heavy.  

\item Chargino decays are indirectly affected via their decays to a lighter neutralino state.  The number of neutralino states lighter than the chargino and their modified compositions alter the chargino branching fractions.  The chargino can decay to an MSSM like $\N_2$ and a singlino $\N_1$, yielding a 5 lepton signal.  Additionally, the extra step in a neutralino decay can allow a 7 lepton final state.  

\item Scenarios exist where the singlet extended models are difficult to differentiate from the MSSM using only the Higgs sector.  However, complementary avenues are available through the discovery of a $Z'$ boson in the UMSSM or extended neutralino cascade decays due to the different neutralino spectrum in singlet extended models.

\end{itemize}

\begin{theacknowledgments}
This work was supported in part by the U.S.~Department of Energy under grant No. DE-FG02-95ER40896, by the Wisconsin Alumni Research Foundation and by the Friends of the IAS.  We thank H-S. Lee for helpful discussions.
\end{theacknowledgments}

\end{document}